\documentclass[conference]{IEEEtran}

\usepackage[dvipdfm]{graphicx}

\usepackage[cmex10]{amsmath}

\ifCLASSINFOpdf
\else
\fi
\usepackage[cmex10]{amsmath}
\usepackage{array}
\begin{document}
%
\title{A Unit Commitment Model with Demand Response for the Integration of Renewable Energies}

\author{
\IEEEauthorblockN{
Yuichi~IKEDA\IEEEauthorrefmark{1},
Takashi~IKEGAMI,
Kazuto~KATAOKA, and 
Kazuhiko~OGIMOTO
}
Collaborative Research Center for Energy Engineering,\\ Institute of Industrial Science, University of Tokyo,\\
4-6-1 Komaba, Meguro-ku, Tokyo 153-8505, JAPAN \\
\IEEEauthorblockA{\IEEEauthorrefmark{1}Email: yikeda@iis.u-tokyo.ac.jp}
}

\maketitle

\begin{abstract}
\boldmath
The output of renewable energy fluctuates significantly depending on weather conditions.
We develop a unit commitment model to analyze requirements of the forecast output and its error for renewable energies.
Our model obtains the time series for the operational state of thermal power plants that would maximize the profits of an electric power utility by taking into account both the forecast of output its error for renewable energies and the demand response of consumers.
We consider a power system consisting of thermal power plants, photovoltaic systems (PV), and wind farms 
and analyze the effect of the forecast error on the operation cost and reserves.
We confirm that the operation cost was increases with the forecast error.
The effect of a sudden decrease in wind power is also analyzed.
More thermal power plants need to be operated to generate power to absorb this sudden decrease in wind power.
The increase in the number of operating thermal power plants within a short period does not affect the total operation cost significantly; 
however the substitution of thermal power plants by wind farms or PV systems is not expected to be very high.
Finally, the effects of the demand response in the case of a sudden decrease in wind power are analyzed.
We confirm that the number of operating thermal power plants is reduced by the demand response.
A power utility has to continue thermal power plants for ensuring supply-demand balance; some of these plants can be decommissioned after installing a large number of wind farms or PV systems, if the demand response is applied using an appropriate price structure.
\end{abstract}



\IEEEpeerreviewmaketitle

\section{Introduction}

To mitigate the effect of climate change, it is essential to reduce the $CO_2$ emissions in various industry sectors by using renewable energies.
The deployment of renewable energies such as wind and photovoltaic (PV) systems has begun in the power industry sector.
The share of renewable energies in the total electricity generation will reach $15\%$ in the near future.
For instance, the climate change package "triple 20" is scheduled to take effect in the European Union; 
the implementation of this package is expected to reduce greenhouse gas emissions by $20\%$ by the year 2020 compared to 1990 levels, cutting consumer energy consumption by $20\%$ by improving energy efficiency, and ensuring that $20\%$ of the EU's energy mix comes from renewables.

It is however noted that the output of renewable energies fluctuates significantly depending on weather conditions.
With an increasing share of energy coming from the renewable energy sources the share of energy contributed by various types of thermal power plants is expected to decrease. 
At the same time, there is an increasing fluctuation in the supply of electric power. 
This implies that the supply-demand balance of electric power cannot be attained using the currently used conventional power systems.

Smart grids have the potential to increase the capability to attain the supply-demand balance of electric power by using new technologies, such as smart meters, home energy management systems, building and energy management systems, control of charge and discharge of electric vehicles, and various types of electric storages \cite{EPRI2008}, \cite{DOE2009}, \cite{Platform2010}, \cite{RoadMap2011}. 
A power utility develops a plan for the operation of a power system based on the forecast output of renewable energies for 24 h. 
The forecast output includes time series of output scenarios and its error for the aggregate outputs of wind power and PV system in the power system under consideration.
An accurate forecast makes it possible to operate the power system economically without making an overestimation of the required adjustments to meet the demand and supply.

From our studies on the operation cost, the $CO_2$ emissions and the peak load of the smart grid \cite{Ikeda2011a}, \cite{Ikeda2011b}, we identified the need to have a  detailed understanding of the time evolution of the power system.        
In this paper, we describe a unit commitment model \cite{Wood1996},\cite{Kerr1966},\cite{Sheble1994},\cite{Padhy2004},\cite{Dang2007},\cite{Salam2007} with demand response for the integration of renewable energies to analyze requirements of the forecast output and its error.
The paper is organized as follows. 
In Section II, the formulation of the model is explained.
In Section III, the analysis of a power system is described.
Finally, Section IV presents our conclusions.

\section{The Model}

The purpose of our unit commitment model is to obtain the time series for the operational state of thermal power plants that would maximize the profit of an electric power utility by taking into account both the forecast of output and its error for renewable energies \cite{Barth2006}, \cite{Wang2009}, \cite{Bu2011}, \cite{Constantinescu2009}, \cite{Monteiro2009} and the demand response of consumers on the change of electricity prices \cite{Lave2007}, \cite{Faruqui2007}, \cite{Faruqui2009b}, \cite{Baia2008}, \cite{Parvania2010}, \cite{Ramchurn2011}.
The model is formulated as a mixed integer linear programming problem. 

\subsection{Objective Function}

The time series of the operational state of thermal power plant $i (i=1,\cdots,N)$ is obtained by maximizing the objective function: 
\begin{equation}
\begin{split}
F(p_t^i,u_t^i,z_t^i,w_t^l) = & \sum_{t=1}^T d_t^{(f)} \sum_{l=1}^{L} w_t^l r^l \Big( \frac{r^l}{\bar{r}} \Big)^{\epsilon_d} \\ 
& - \sum_{t=1}^T \sum_{i=1}^N [b_i p_t^i + S_i z_t^i].
\label{eq:ObjFnc}
\end{split}
\end{equation}
This objective function represents the profit of an electric power utility.
The first term of the r.h.s. in Eq. (\ref{eq:ObjFnc}) is the sales revenue and the second term is the operation cost.
Here $N$, $T$, $L$ are the number of thermal power plants, time horizon, and the number of price levels, respectively.
Continuous variables $p_t^i$ is the output power variable of thermal power plant $i$, and integer variables $u_t^i$, $z_t^i$, $w_t^l$ are	
the status production variable of thermal power plant $i$ (1=committed, 0=decommitted),
the start-up variable of thermal power plant $i$ (1=start up, 0=others), and 
the demand response variable (1=selected, 0=not selected), respectively.
Parameters $S_i$ and $b_i$are the start-up cost of thermal power plant $i$ and the fuel cost of the thermal power plant $i$, respectively. 
The forecasted demand and its error are indicated by $d_t^{(f)}$ and $\sigma_d$, respectively. Here $(f)$ stands forforecasting.

\begin{figure}
\begin{center}
\includegraphics[scale=0.5]{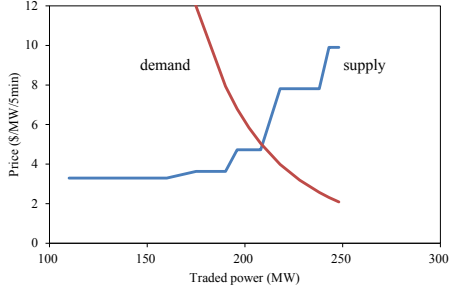}
\caption{Demand and supply}
\label{fig:ds}
\end{center}
\end{figure}

Other parameters related to the demand response $\bar{r}$, $r^l$, $\epsilon_d$ are the average electricity price, the price level, and the price elasticity of demand, respectively. 
If the electricity price $r$ deviates from the average price $\bar{r}$, the demand $d$ is changed from the average demand $\bar{d}$ as follows:
\begin{equation}
\frac{d}{\bar{d}} = \Big( \frac{r}{\bar{r}} \Big)^{\epsilon_d}.
\label{eq:Dcurve}
\end{equation}
The dependence of the demand $d$ on the price $r$ is depicted in Fig. \ref{fig:ds} \cite{Arroyo2000}, \cite{Torre2002}, \cite{Borghetti2003}, \cite{Ventosa2005}, \cite{Weidlich2008}.

\subsection{Global Constraints}

The sum of the demand response variable $w_t^l$ has to satisfy the constraint
\begin{equation}
\sum_{l=1}^{L} w_t^l = 1
\label{eq:Plevel}
\end{equation}
to ensure that only a single price level $r^l$ is selected.
In addition to this constraint, the average of the selected price $r^l$ has to be equal to the average price $\bar{r}$
\begin{equation}
\frac{1}{T} \sum_{t=1}^T \sum_{l=1}^{L} w_t^l r^l \le \bar{r}.
\label{eq:meanP}
\end{equation}
Moreover the total demand has to be unchanged by the demand response:
\begin{equation}
\sum_{t=1}^T d_t^{(f)} \le \sum_{t=1}^T d_t^{(f)} \sum_{l=1}^{L} w_t^l \Big( \frac{r^l}{\bar{r}} \Big)^{\epsilon_d}. 
\label{eq:Dconserve}
\end{equation}
Importantly, the sum of supply has to be greater than the demand:
\begin{equation}
\sum_{i=1}^N p_t^i+wd_t^{(f)}+pv_t^{(f)} \ge d_t^{(f)}\sum_{l=1}^{L}w_t^l \Big( \frac{r^l}{\bar{r}} \Big)^{\epsilon_d},
\label{eq:DSbalanceNoSigma1}
\end{equation}
where $wd_t^{(f)}$ and $pv_t^{(f)}$ are the forecasted wind power generation and the forecasted PV generation, respectively.

\begin{figure}
\begin{center}
\includegraphics[scale=0.5]{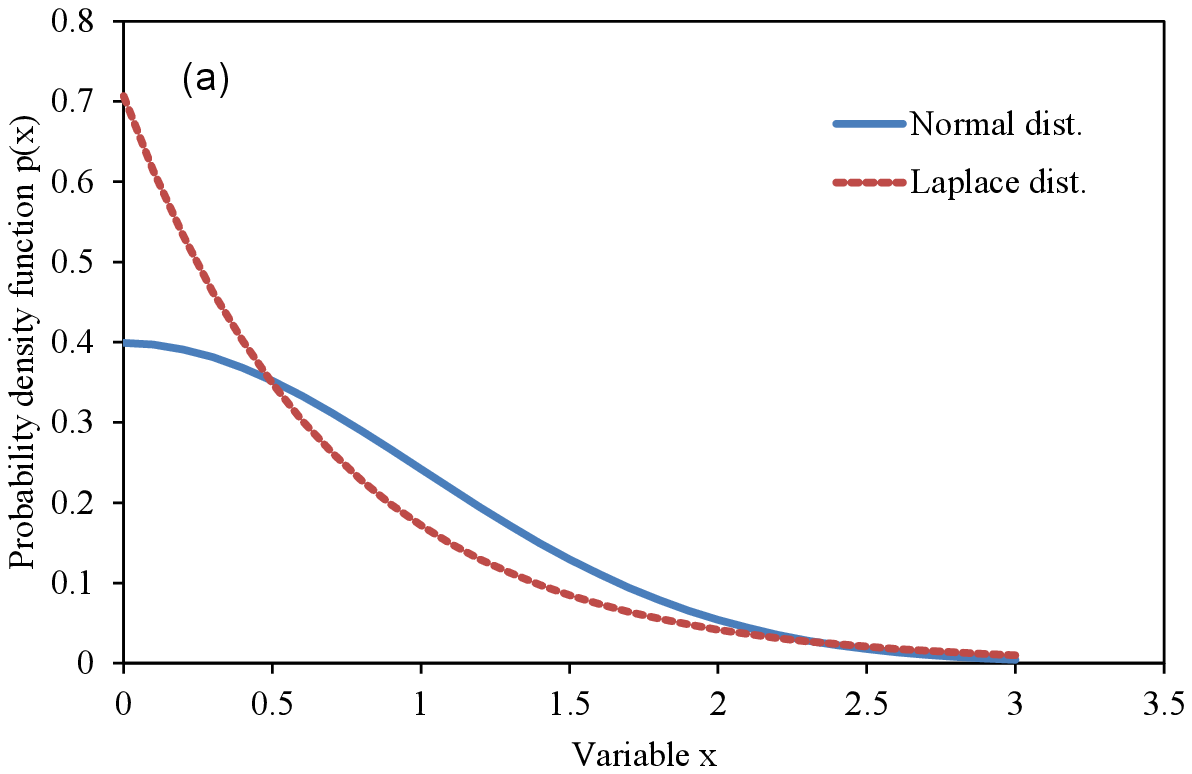}
\includegraphics[scale=0.5]{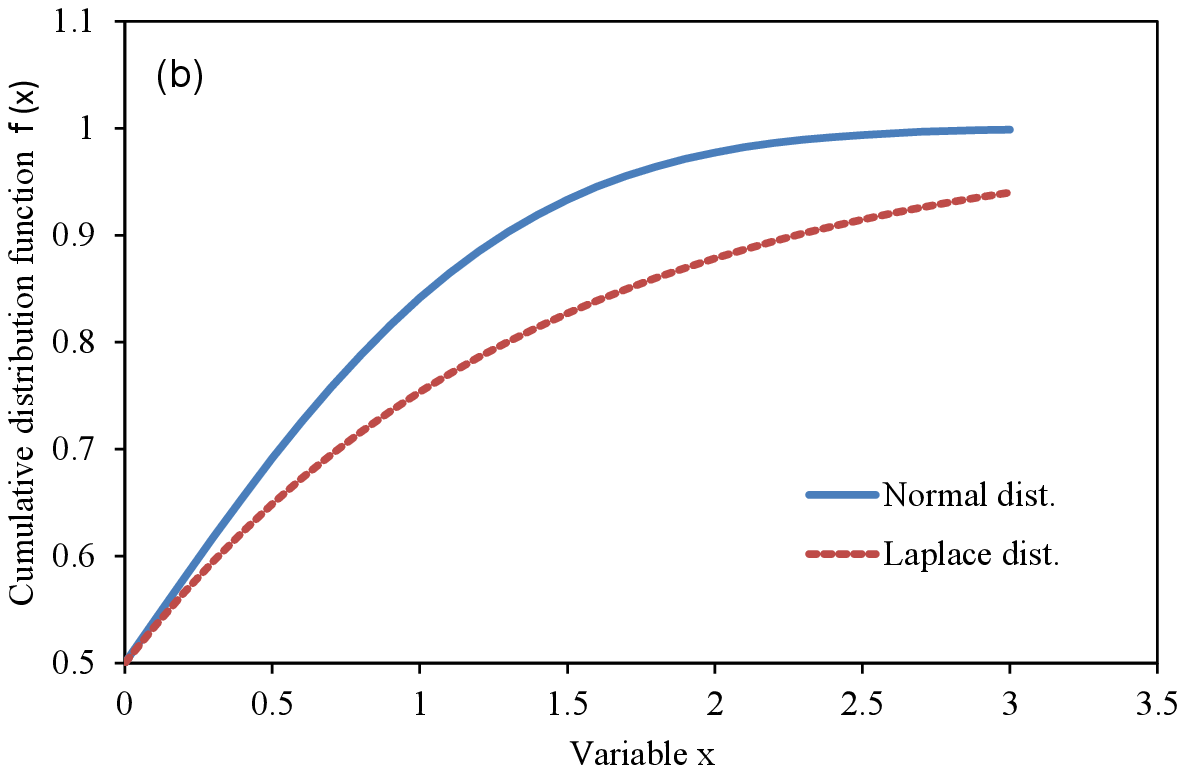}
\caption{Distribution of forecast error}
\label{fig:dist}
\end{center}
\end{figure}

If we consider the forecast error of demand $\sigma_d$, the forecast error of wind power $\sigma_w$, and the forecast error of PV $\sigma_p$,
the constraint in Eq.(\ref{eq:DSbalanceNoSigma1}) can be rewritten as
\begin{equation}
\frac{\sum_{i=1}^N p_t^i+wd_t^{(f)}+pv_t^{(f)}-d_t^{(f)}\sum_{l=1}^{L}w_t^l \Big( \frac{r^l}{\bar{r}} \Big)^{\epsilon_d}}{\sqrt{\sigma_d^2+\sigma_w^2+\sigma_p^2}} \ge \phi^{-1}(\alpha),
\label{eq:DSbalance}
\end{equation}
where $\alpha$ and $\phi(\cdot)$ are the probability to ensure the supply-demand balance and the cumulative distribution function, respectively.
If the forecast error is distributed according to the normal distribution, the probability density function is
\begin{equation}
p(x) = \frac{1}{\sqrt{2 \pi \sigma^2}} exp \Big[-\frac{(x-\mu)^2}{2 \sigma^2} \Big],
\label{eq:NormalPDF}
\end{equation}
and the cumulative distribution function is written using the error function $erf[\cdot]$ as 
\begin{equation}
\phi(x) = \frac{1}{2} \Big( 1 + erf \Big[\frac{x-\mu}{\sqrt{2 \sigma^2}} \Big] \Big),
\label{eq:NormalCDF}
\end{equation}
where $\mu$ and $\sigma$ are a mean and a standard deviation, respectively.
However, if the probability density function $p(x)$ is a Laplace distribution
\begin{equation}
p(x) = \frac{1}{2b} exp \Big[-\frac{|x-\mu|}{b} \Big],
\label{eq:LaplacePDF}
\end{equation}
then the cumulative distribution function $\phi(x)$ is
\begin{equation}
\phi(x) = \frac{1}{2} \Big( 1 + sgn(x-\mu) \Big(1 - exp \Big[-\frac{|x-\mu|}{b} \Big] \Big) \Big).
\label{eq:LaplaceCDF}
\end{equation}
Here a standard deviation is given by $\sigma=\sqrt{2}b$ and $sgn(x-\mu)=+(x\ge\mu), -(x<\mu)$.
The functional forms for these distributions are depicted for $\mu=0$ and $\sigma=1$ in Fig. \ref{fig:dist}.
It is noted that here the Laplace distribution shows a distribution tail longer than the normal distribution.
This implies that the probability to ensure supply-demand balance differs in these two distributions.

\subsection{Local Constraints for Thermal Power Plants}

The following constraints are used for each thermal power plant as typical constraints in a unit commitment model.

\subsubsection{Generation Capacity}

The output power $p_t^i$ has to be between the maximum output power $\bar{p}_{max}^i$ and the minimum output power $\bar{p}_{min}^i$ when the operation is in steady state:
\begin{equation}
\bar{p}_{min}^i u_t^i \le p_t^i \le \bar{p}_{max}^i u_t^i.
\label{eq:Capa}
\end{equation}

\subsubsection{Ramp-up Limit}

The increase in the output of thermal power plant $i$ has to be smaller than the maximum ramp-up speed $\Delta_+$ when the unit is up at the previous time step
and is smaller than the minimum output power $\bar{p}_{min}^i$ when the unit is down at the previous time step:
\begin{equation}
p_t^i - p_{t-1}^i \le u_{t-1}^i \Delta_+^i + (1-u_{t-1}^i) \bar{p}_{min}^i.
\label{eq:RumpUp}
\end{equation}

\subsubsection{Ramp-down Limit}

The decrease in the output of thermal power plant $i$ has to be smaller than the maximum ramp-down speed $\Delta_-$ when the unit is up at time step $t$
and is smaller than the maximum output power $\bar{p}_{max}^i$ when the unit is down at time step $t$:
\begin{equation}
p_t^i - p_{t-1}^i \ge - u_t^i \Delta_-^i - (1-u_t^i) \bar{p}_{max}^i.
\label{eq:RumpDown}
\end{equation}

\subsubsection{Minimum Up-time Constraint}

Thermal power plant $i$ has to be operated longer than the minimum up-time requirement $\tau_+^i$, once the unit is up:
\begin{gather}
u_t^i \ge u_s^i - u_{s-1}^i, \notag \\
s \in [t-\tau_+^i, t-1].  
\label{eq:MinUp}
\end{gather}

\subsubsection{Minimum Down-time Constraint}

Thermal power plant $i$ has to be stopped longer than the minimum down-time requirement $\tau_-^i$, once the unit is down:
\begin{gather}
u_t^i \le 1 + u_s^i - u_{s-1}^i, \notag \\
s \in [t-\tau_-^i, t-1]. 
\label{eq:MinDown}
\end{gather}

\subsubsection{Constraint on the Start-up Variable}

The start-up variable $z_t^i$ has to satisfy the following constraints by definition:
\begin{gather}
z_1^i \ge u_1^i,  \notag \\
z_t^i \ge u_t^i - u_{t-1}^i (t>2).
\label{eq:Start}
\end{gather}

\section{Analyzed Power System}
\label{sec:SamllSystem}

\begin{figure}
\begin{center}
\includegraphics[scale=0.5]{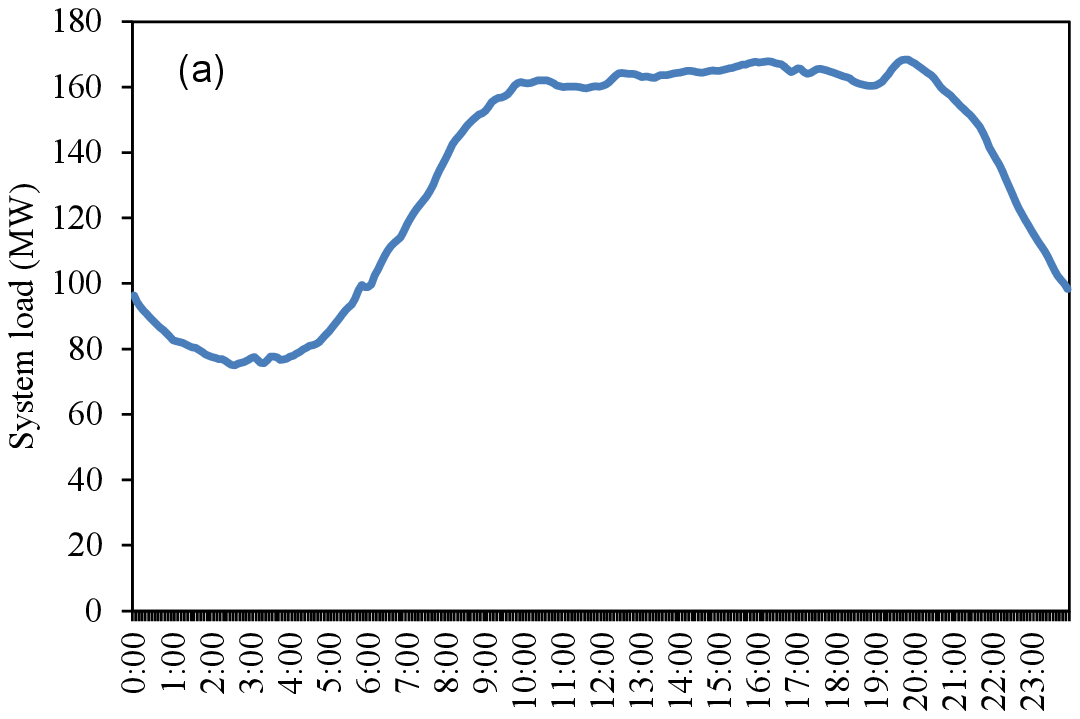}
\includegraphics[scale=0.5]{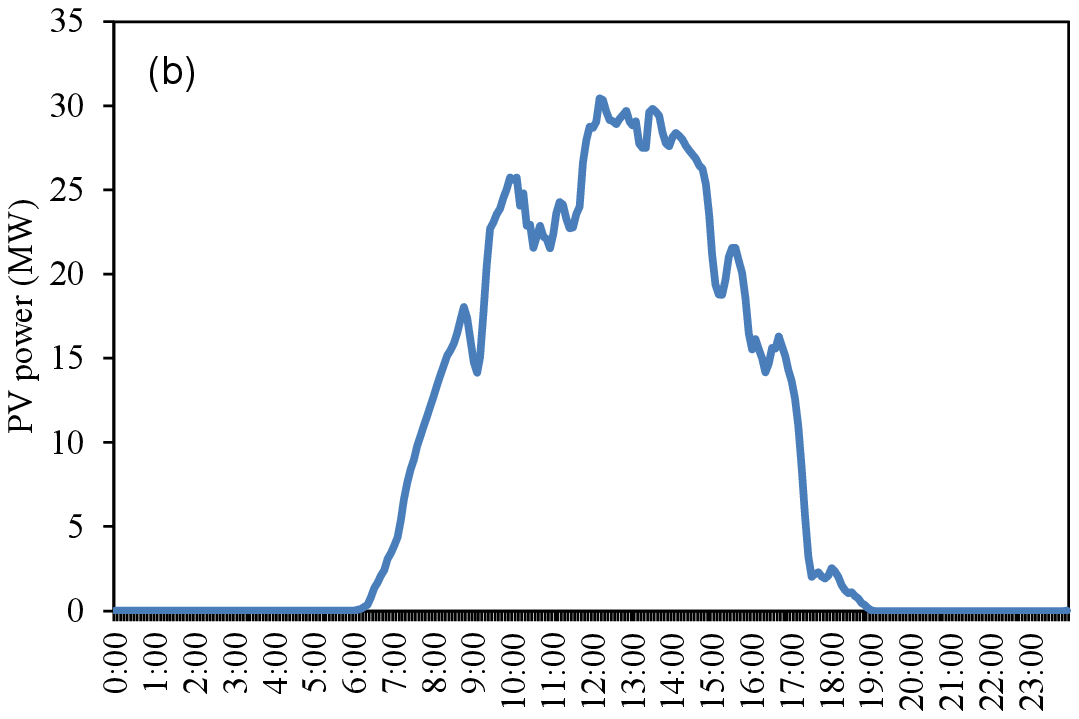}
\includegraphics[scale=0.5]{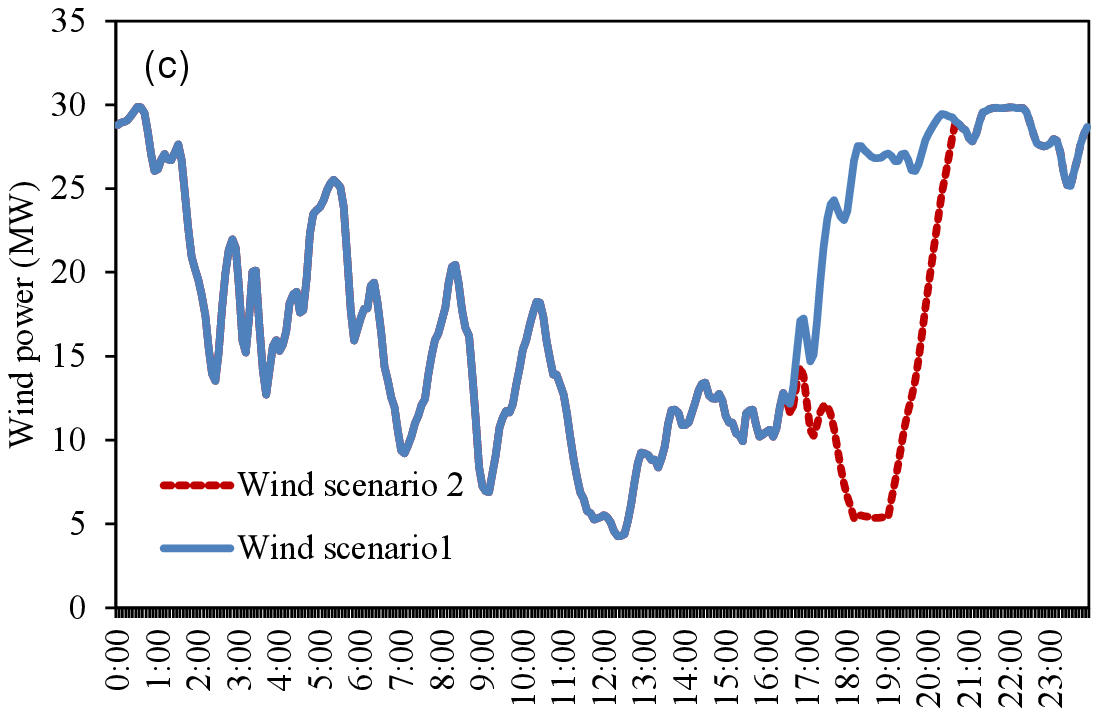}
\caption{Given load, PV, wind scenarios}
\label{fig:givensenario}
\end{center}
\end{figure}

We analyze the time series for the operational state of thermal power plants by taking into account both forecast output and its error for renewable energies and the demand response of consumers.
We consider a small power system consisting of thermal power plants, PV systems, and wind farms in this study, although the model is easily extendible to a larger power system.
The number of thermal power plants is $12$, and parameters are given in Table \ref{ThermUnit}.
The installed capacity for each PV system and wind farm is $30 MW$.
The intra-day peak demand is about $170MW$.
We assume the forecast errors $\sigma_w$ and $\sigma_p$ are both $10\% (3MW)$ of the installed capacity in the reference case.
In addition to the reference case, we analyze two more cases $\sigma_w=6MW$ and $\sigma_w=9MW$, while $\sigma_p$ remains unchanged.
The forecast error for the demand is $\sigma_d=0$ for all cases.
Scenarios for the demand, the PV output power, and the wind output power in the reference case are shown in Fig. \ref{fig:givensenario}.
While wind scenario 1 is used in the reference case, wind scenario 2, where wind power suddenly decreases in the evening, is used to discuss the fluctuation of wind output and the effect on the demand response. 

\begin{table}[!t]
\renewcommand{\arraystretch}{1.3}
\caption{The parameters of thermal power plants}
\label{ThermUnit}
\centering
\begin{tabular}{c|c c c c c c c c}
\hline
$i$ & $b_i$ & $\bar{p}_{max}^i$ & $\bar{p}_{min}^i$ & $\Delta_+$ & $\Delta_-$ & $\tau_+^i$ & $\tau_-^i$ & $S_i$ \\
\hline
\hline
1 & 3.0 & 50.0 & 25.0 & 0.5 & 0.5 & 3.0 & 3.0 & 1000.0 \\
2 & 3.0 & 50.0 & 25.0 & 0.5 & 0.5 & 3.0 & 3.0 & 1000.0 \\
3 & 3.3 & 15.0 & 7.5 & 0.5 & 0.5 & 3.0 & 3.0 & 200.0 \\
4 & 3.3 & 15.0 & 7.5 & 0.5 & 0.5 & 3.0 & 3.0 & 200.0 \\
5 & 4.3 & 6.0 & 2.0 & 5.0 & 5.0 & 3.0 & 3.0 & 100.0 \\
6 & 4.3 & 6.0 & 2.0 & 5.0 & 5.0 & 3.0 & 3.0 & 100.0 \\
7 & 4.3 & 6.0 & 2.0 & 5.0 & 5.0 & 3.0 & 3.0 & 100.0 \\
8 & 7.1 & 10.0 & 4.0 & 5.0 & 5.0 & 3.0 & 3.0 & 200.0 \\
9 & 7.1 & 10.0 & 4.0 & 5.0 & 5.0 & 3.0 & 3.0 & 200.0 \\
10 & 7.1 & 10.0 & 4.0 & 5.0 & 5.0 & 3.0 & 3.0 & 200.0 \\
11 & 9.0 & 5.0 & 2.5 & 0.5 & 0.5 & 3.0 & 3.0 & 100.0 \\
12 & 9.0 & 5.0 & 2.5 & 0.5 & 0.5 & 3.0 & 3.0 & 100.0 \\
\hline
\end{tabular}
\end{table}

The marginal cost $C_t^M$, the spinning reserve $R_t^S$, and the load frequency control (LFC) margin $R_t^L$ are calculated using the obtained time series $p_t^i (i=1,\cdots,N)$.
The marginal cost $C_t^M$ is defined by
\begin{equation}
C_t^M = \max \{ b_i | u_t^i=1 \}.
\label{eq:Mcost}
\end{equation}
The spinning reserve $R_t^S$ and the LFC margin $R_t^L$ are defined as proxy quantities in this study as follows:
\begin{equation}
R_t^S = \sum_{i=1}^N p_t^i - \sum_{i=1}^N \hat{p}_t^i,
\label{eq:SpinRes}
\end{equation}
\begin{equation}
R_t^L = \sum_{i=1}^N \min \{ \bar{p}_{max}^i - p_t^i, 0.05 \bar{p}_{max}^i | u_t^i=1 \}.
\label{eq:LFCRes}
\end{equation}
Here, $\hat{p}_t^i$ is the output power of thermal power plant $i$ obtained in the optimization with the constraint of Eq.(\ref{eq:DSbalanceNoSigma1}).
The spinning reserve $R_t^S$ is ready to generate power to absorb the fluctuation instantaneously.
On the other hand, the LFC margin $R_t^L$ is the remaining capacity that is able to increase output with the constraints of Eq. (\ref{eq:RumpUp}) and Eq. (\ref{eq:RumpDown}).

\section{Results}

We discuss the requirements on the forecast output and its error for the integration of the renewable energies in this section.
The unit commitment model formulated as a mixed integer linear programming problem in Eqs. (\ref{eq:ObjFnc}) to (\ref{eq:Start}) was solved to analyze the power system described in Section \ref{sec:SamllSystem} using a commercial solver \cite{Frangioni2008}, \cite{Frangioni2009}. 

\subsection{Reference Case}

\begin{figure}
\begin{center}
\includegraphics[scale=0.5]{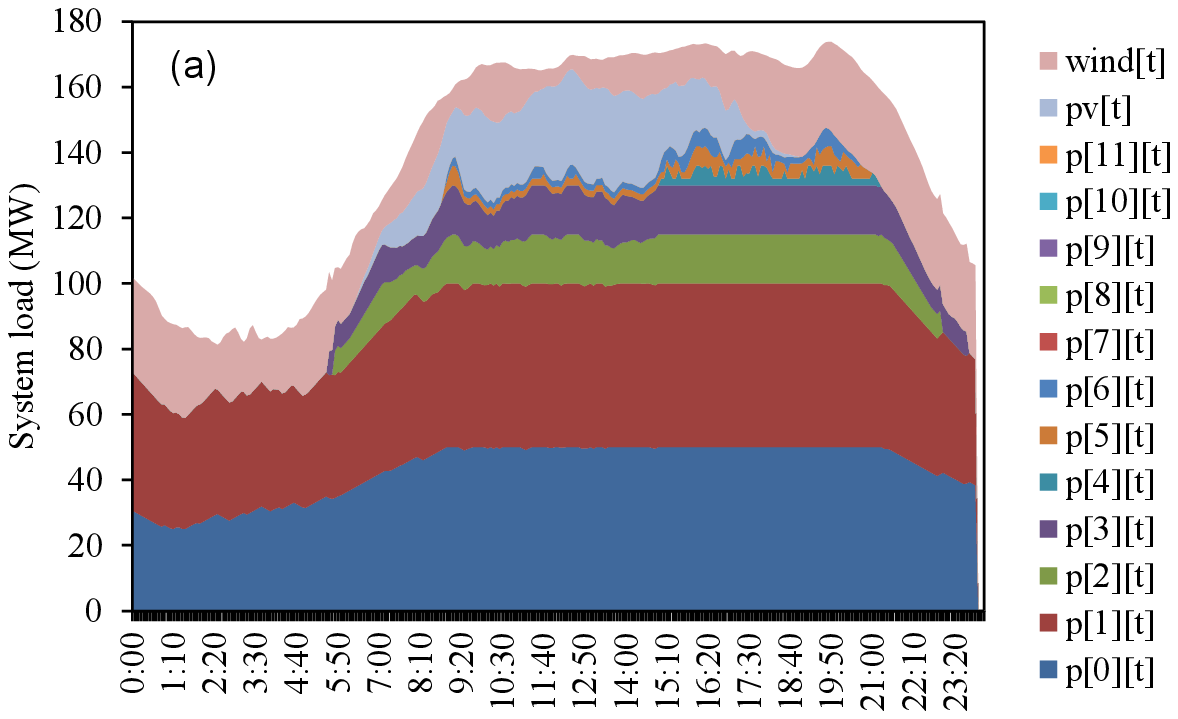}
\includegraphics[scale=0.5]{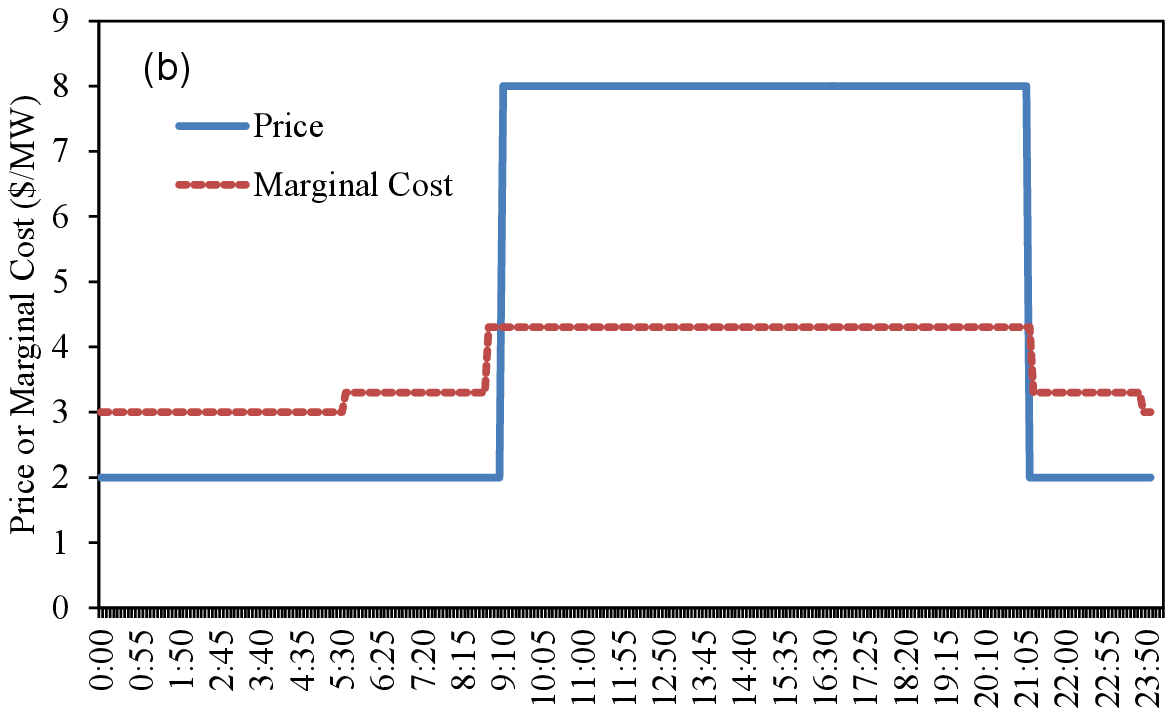}
\includegraphics[scale=0.5]{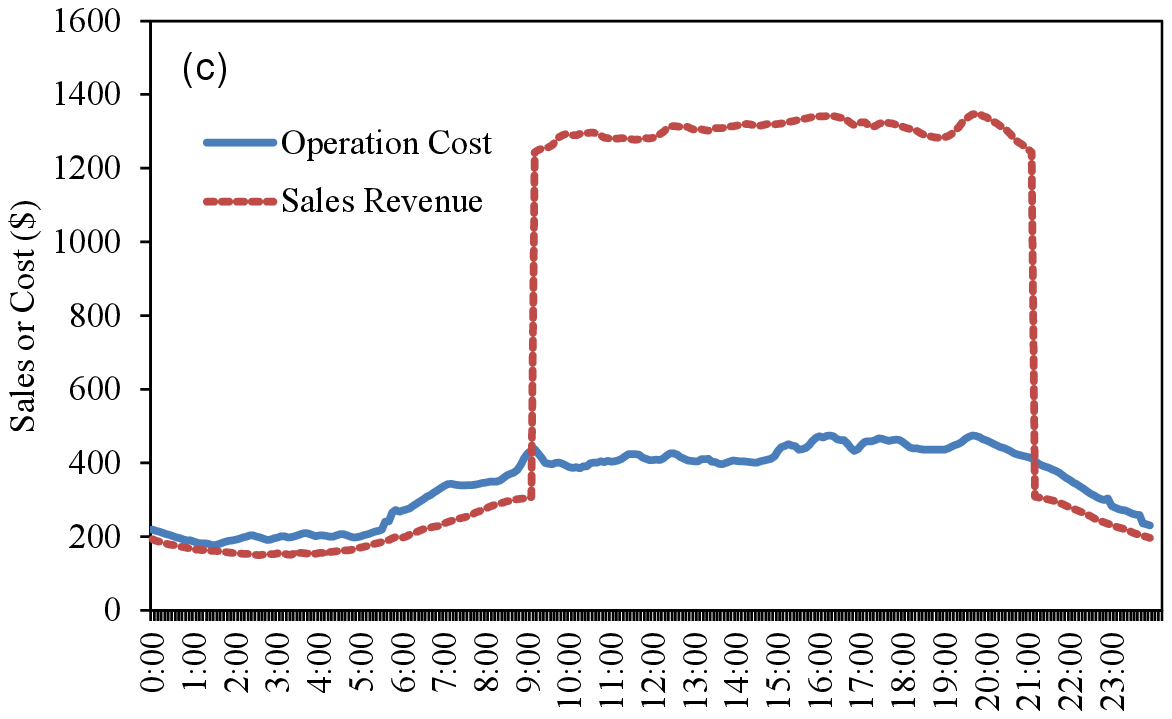}
\includegraphics[scale=0.5]{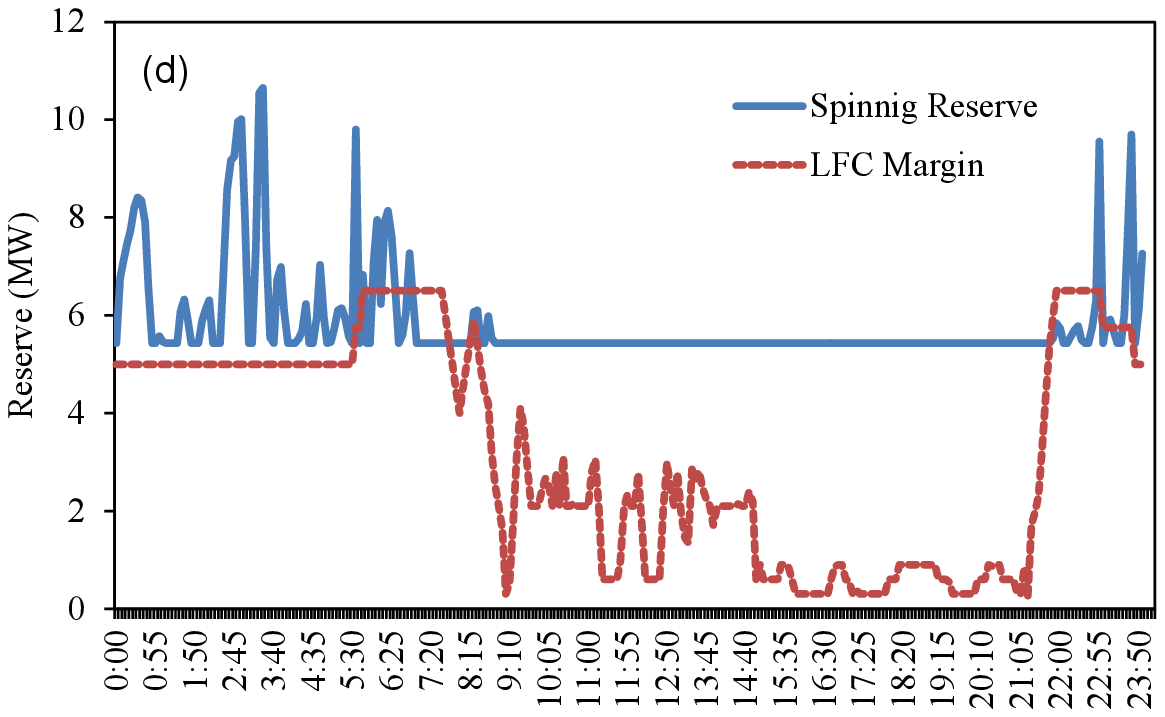}
\caption{Results of the reference case}
\label{fig:reference}
\end{center}
\end{figure}

The results of the reference case are shown in Fig. \ref{fig:reference}.
Figure \ref{fig:reference} (a) depicts the share of the thermal power plants, the wind power, and the PV systems to satisfy the given demand.
The thermal power plants are numbered in the order of increasing operation cost, i.e., the merit order. 
The fluctuation of the wind and the PV outputs was absorbed by starting the thermal power plants serially in the merit order.  
Figure \ref{fig:reference} (b) shows that the marginal cost $C_t^M$ is high from $9:00$ to $20:00$. This is because some additional thermal power plants are operating in this period.
The price structure where the daytime rate is higher than the nighttime rate is obtained as a result of maximizing the profits of an electric power utility. 
In a reflection of the price structure, the sales revenue is high during daytime as shown in Fig. \ref{fig:reference} (c).
Figure \ref{fig:reference} (d) depicts that the spinning reserve $R_t^S$ has a flat structure during daytime, while the LFC margin $R_t^L$ is high during nighttime.

\subsection{Demand Response}

\begin{figure}
\begin{center}
\includegraphics[scale=0.5]{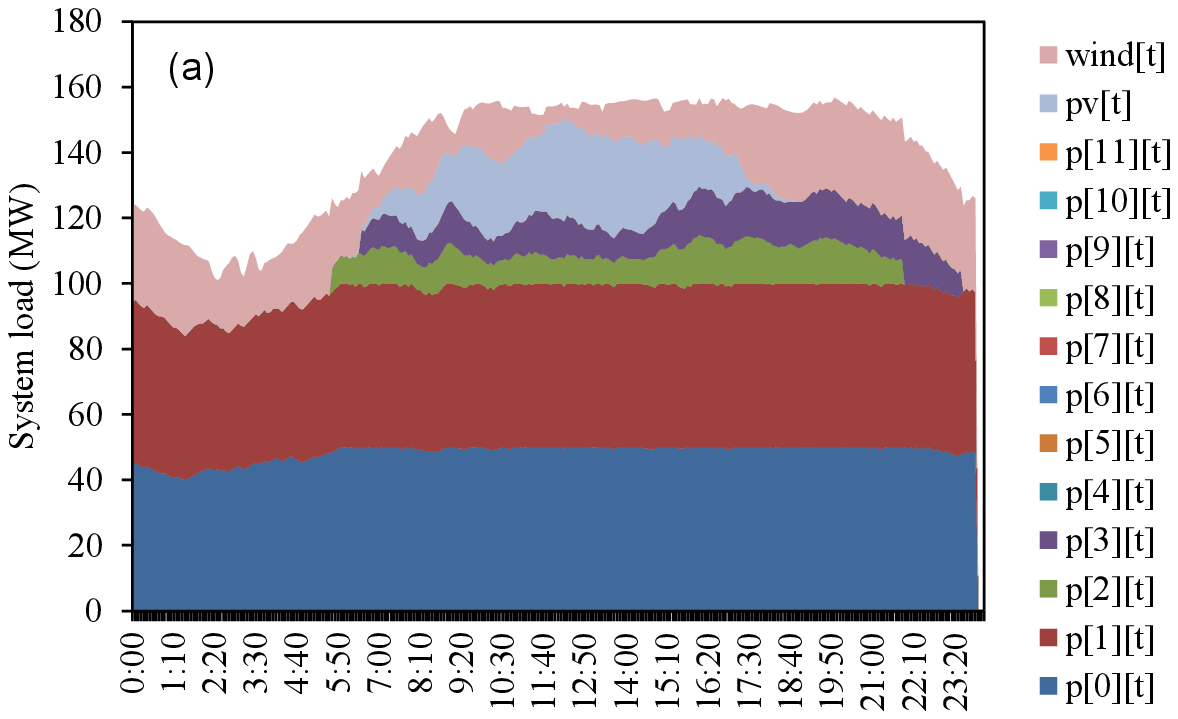}
\includegraphics[scale=0.5]{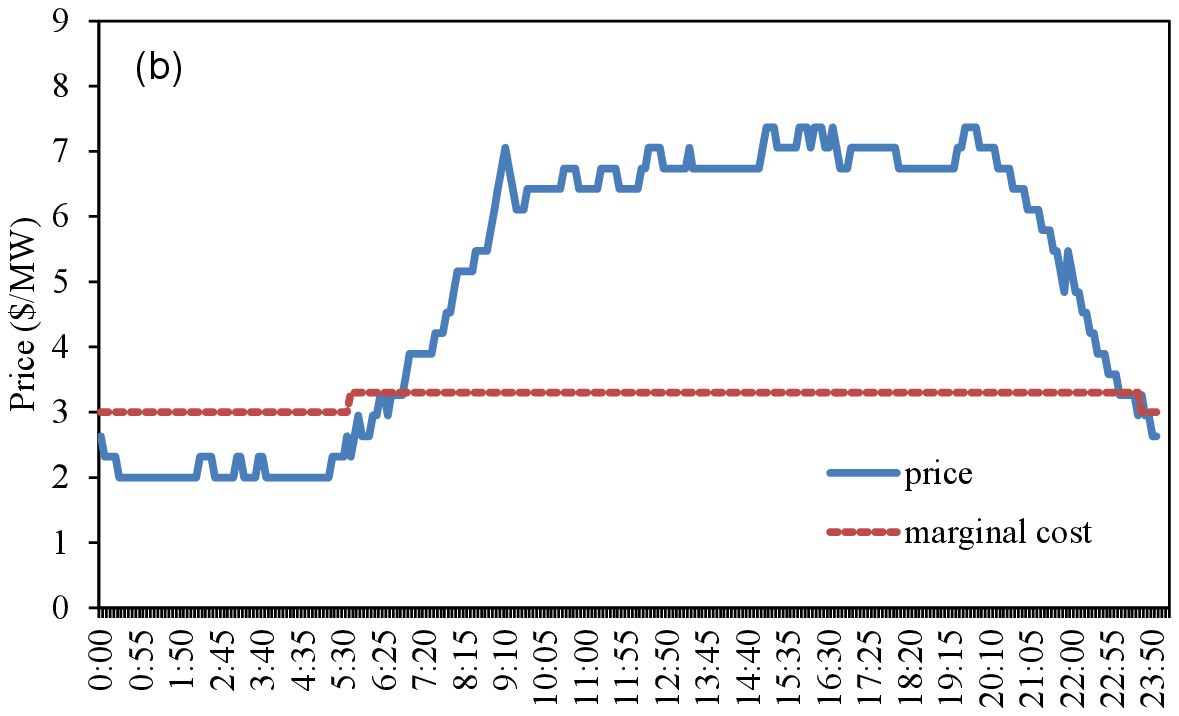}
\caption{Demand response}
\label{fig:dr}
\end{center}
\end{figure}

It is well known that the price elasticity of demand $\epsilon_d$ is small because the share of expenditure for electricity in a household budget is low \cite{Kirschen2003}.
The effects of the demand response with $\epsilon_d=-0.30$ are shown in Fig. \ref{fig:dr}. 
Figure \ref{fig:dr} (a) depicts that the load profile becomes flat and the peak load is reduced by about $20MW$.
The profile of the marginal cost becomes flat and the price structure is significantly changed as shown in Fig. \ref{fig:dr} (b).
In general, a larger effect to the demand response is expected for a larger $\epsilon_d$.

\subsection{Effects of the Forecast Error}

\begin{figure}
\begin{center}
\includegraphics[scale=0.5]{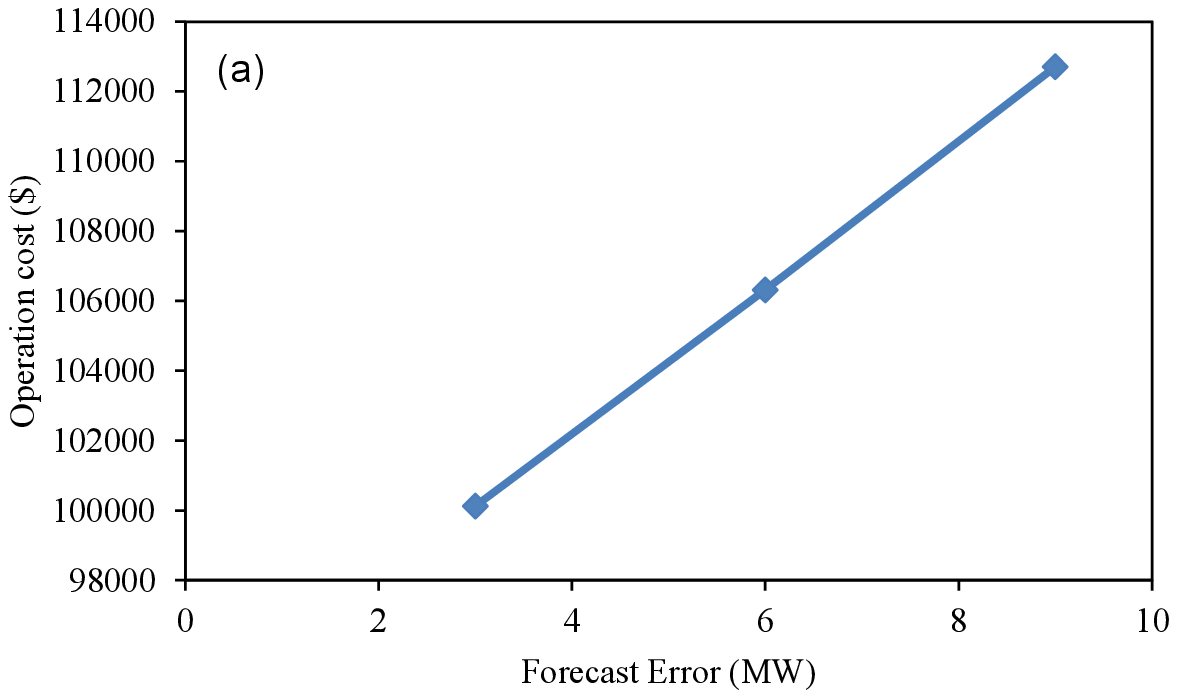}
\includegraphics[scale=0.5]{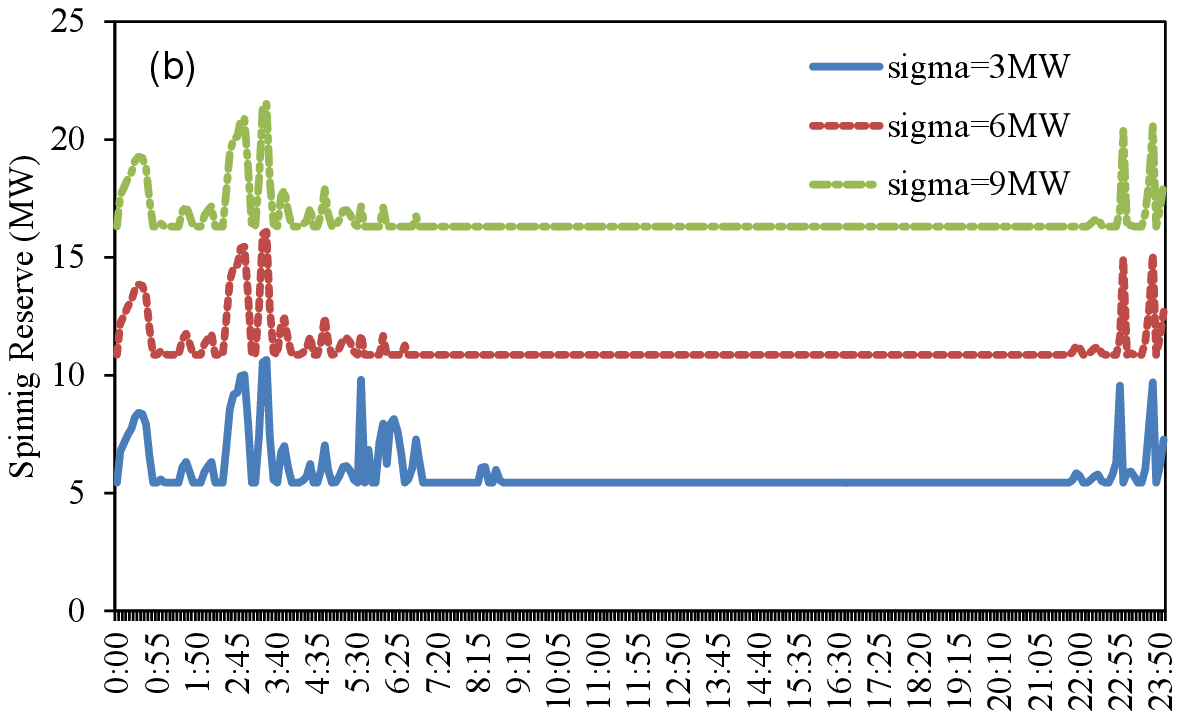}
\includegraphics[scale=0.5]{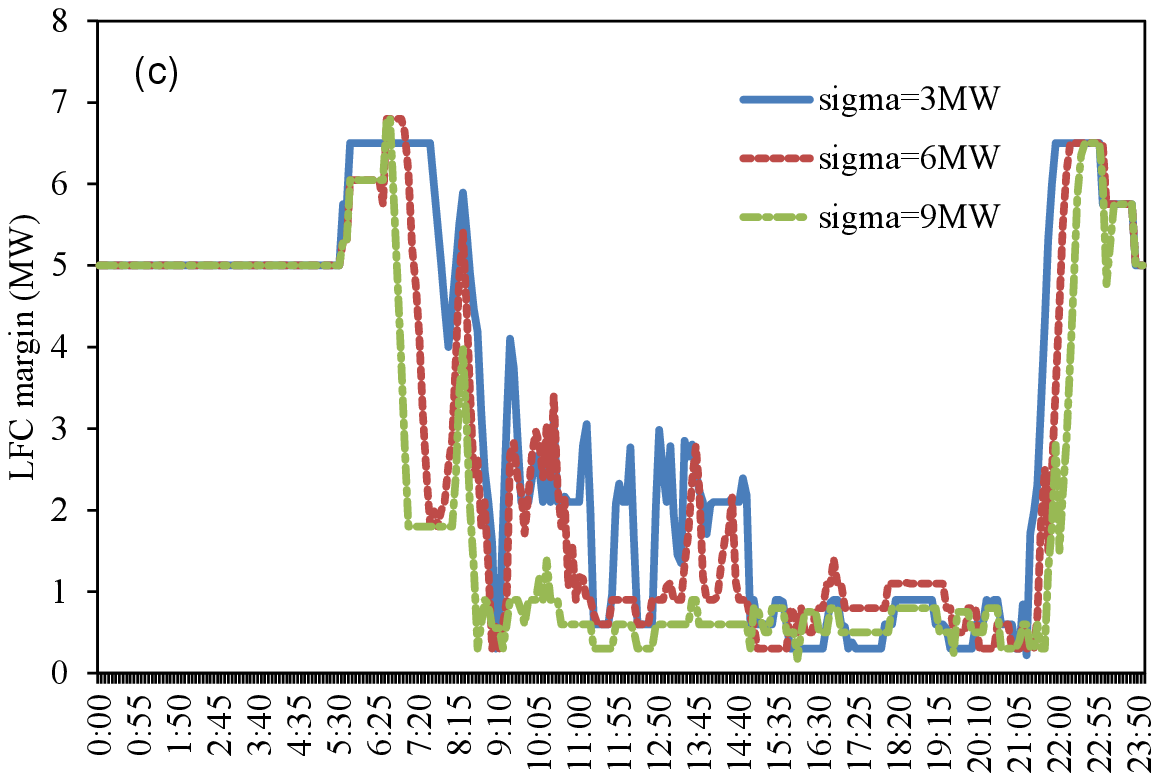}
\caption{Effect of forecast error on cost and reserves}
\label{fig:effecterror}
\end{center}
\end{figure}

The effects of the forecast error on the operation cost and reserves are analyzed here.
The forecast errors $\sigma_w$ and $\sigma_p$ are $10\% (3MW)$ of the installed capacity in the reference case.
In addition to the reference case we analyzed two more cases; $\sigma_w=\sigma_p=6MW$ and $\sigma_w=\sigma_p=9MW$.
Figure \ref{fig:effecterror} (a) depicts that the operation cost increases with the forecast error $\sigma_w$.
This tendency is also true for the spinning reserve $R_t^S$ as shown in Fig. \ref{fig:effecterror} (b).
The intra-day structure is the same for the three cases, but the level increases by $5.4 MW$ as $\sigma_w$ increases by $3 MW$. 
This is reasonable because a larger reserve is required to absorb the larger fluctuation.
Figure \ref{fig:effecterror} (c) shows that the LFC margin $R_t^L$ decreases during daytime, while the margin for all three were relatively high during nighttime, independently of the forecast error.
This implies that the demand during nightime is relatively low, and therefore, the remaining capacity is large in the operating thermal power plants.

\subsection{Effects of Sudden Decrease in Wind Power}

\begin{figure}
\begin{center}
\includegraphics[scale=0.5]{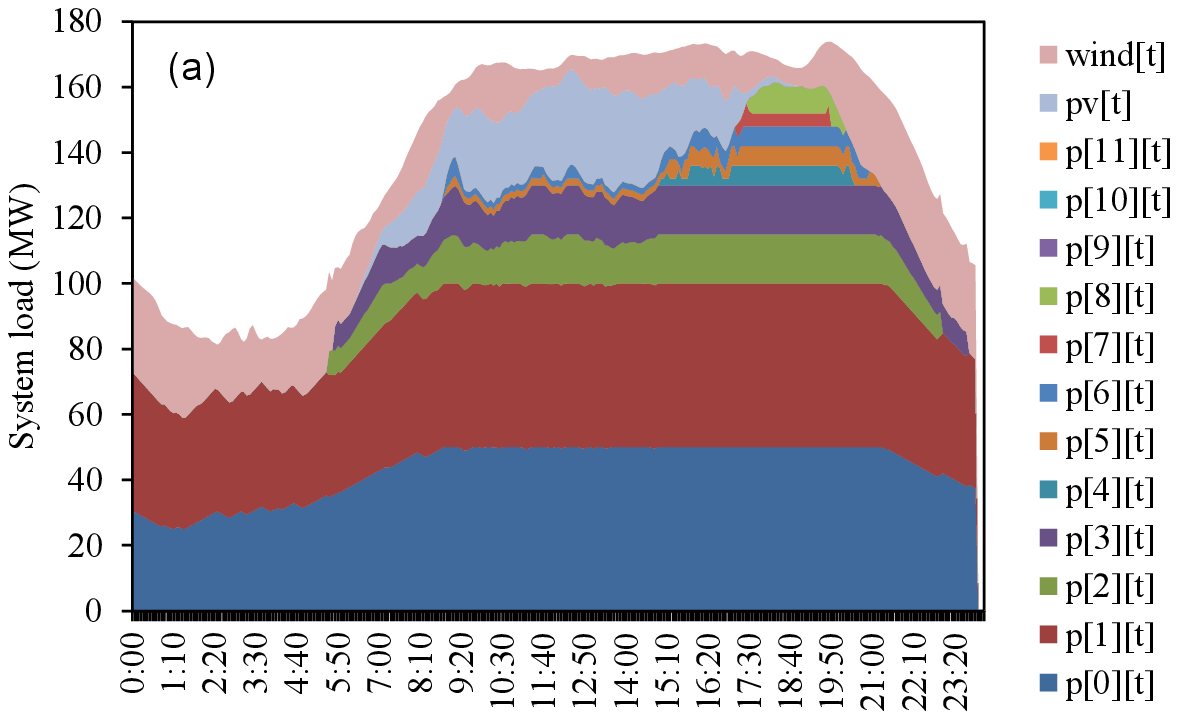}
\includegraphics[scale=0.5]{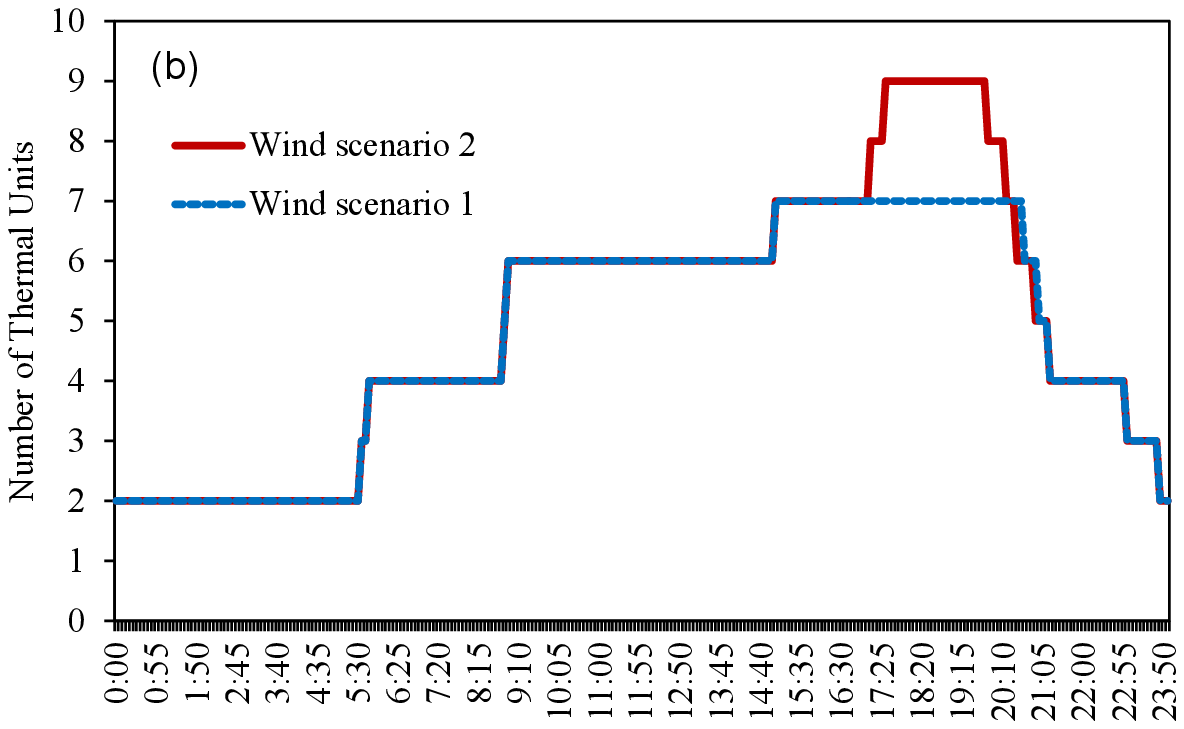}
\caption{Wind scenario 2}
\label{fig:w2}
\end{center}
\end{figure}

In wind scenario 2, the wind power suddenly decreased in the evening compared to wind scenario 1.
We expect that more thermal power plants will be operated to generate power to absorb this sudden decrease in wind power.
The effects of the sudden decrease in wind power were analyzed and the results are shown in Fig. \ref{fig:w2}.
Figure \ref{fig:w2} (a) depicts an increase in the number of operating thermal power plants between $17:00$ and $20:00$.
This increase is more clearly shown in Fig. \ref{fig:w2} (b). Nine thermal power plants are in operation during $17:00$ and $20:00$ for scenario 2,
whereas just seven plants operate in the same period for the scenario 1.
The increase in the number of operating thermal power plants in such a short period does not affect the total operation cost significantly.
It is however noted that the power utility has to continue to use thermal power plants for ensuring supply-demand balance even after installing a large number of wind farms or PV systems.
Therefore, the substitution of thermal power plants by wind farms or PV systems is not expected to be very high, although this issue has to be studied quantitatively using actual data.

\subsection{Demand Response and a Sudden Decrease in Wind Power}

\begin{figure}
\begin{center}
\includegraphics[scale=0.5]{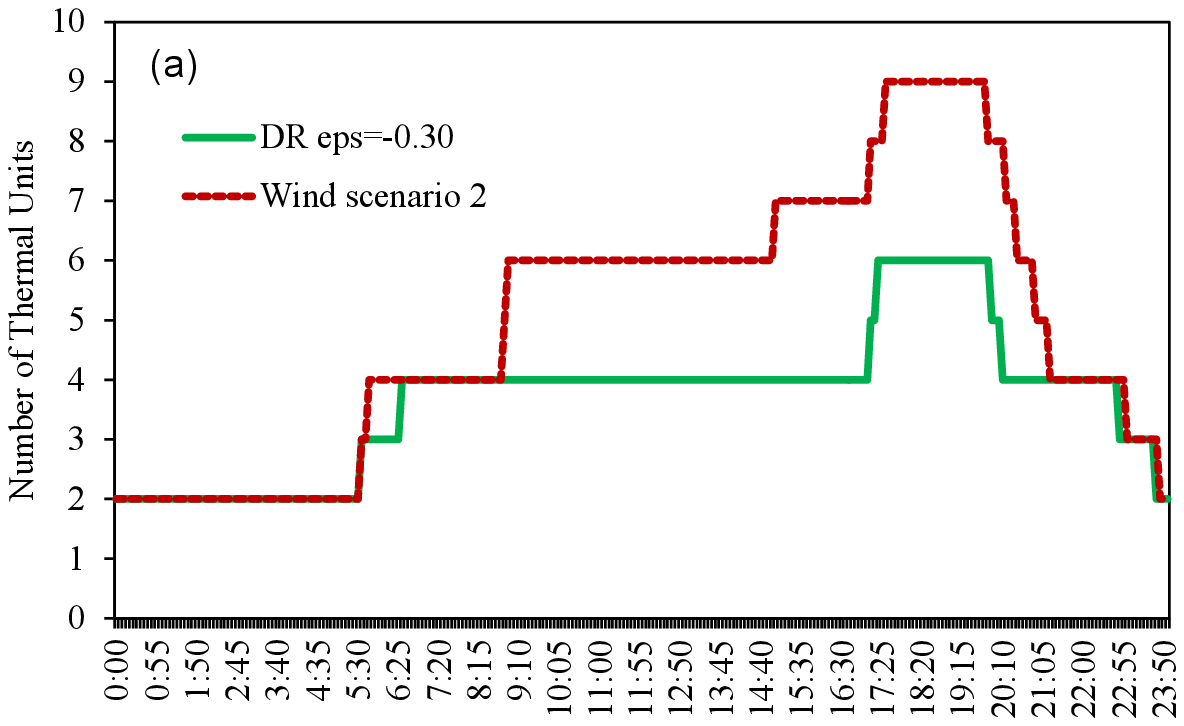}
\includegraphics[scale=0.5]{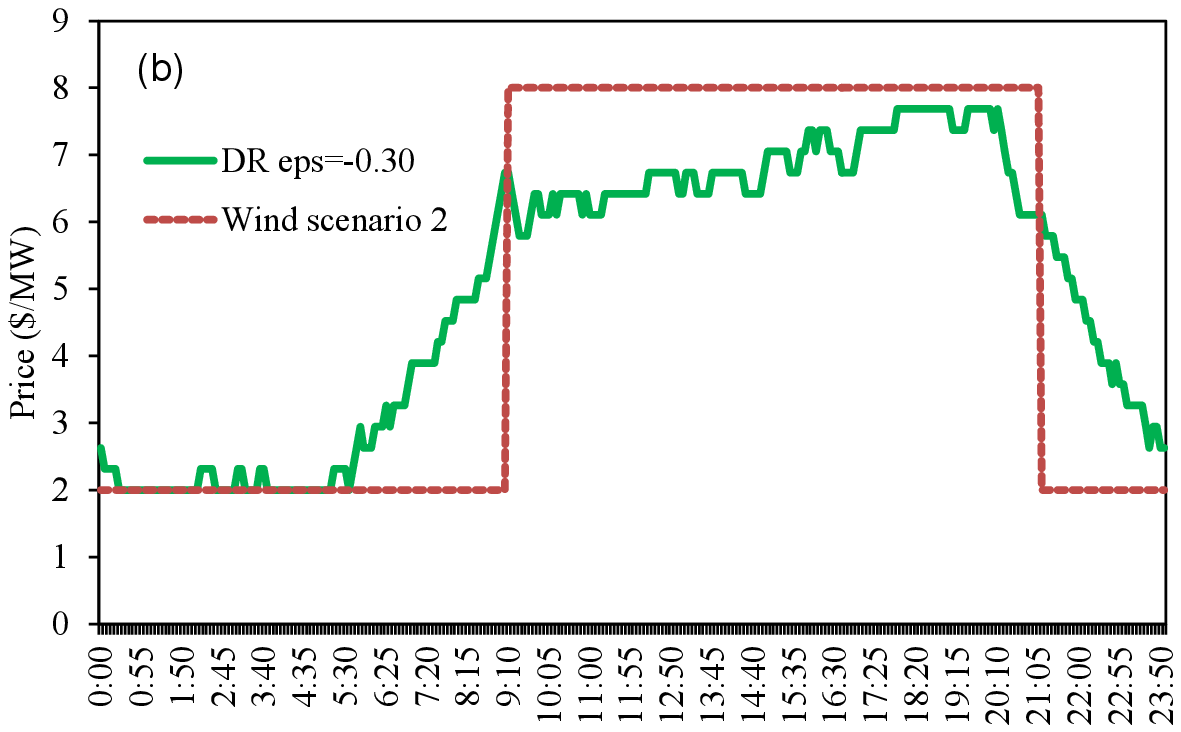}
\includegraphics[scale=0.5]{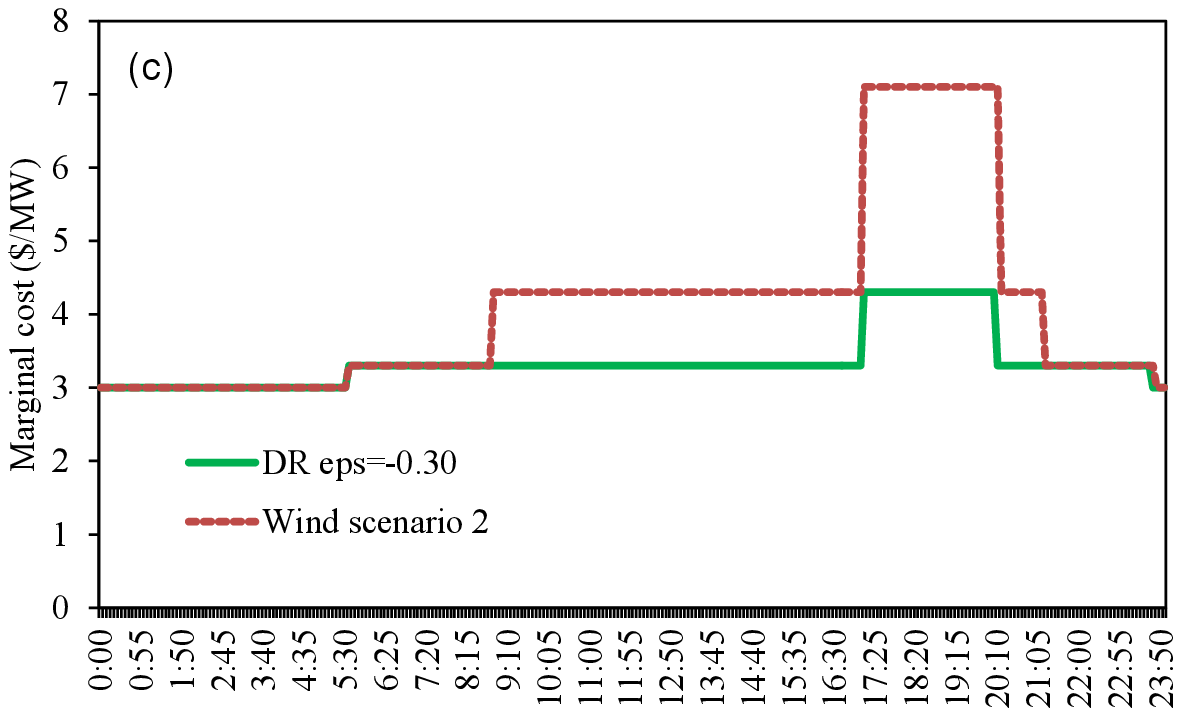}
\caption{Wind scenario 2 with demand response}
\label{fig:w2dr}
\end{center}
\end{figure}

Finally, the effects of the demand response with $\epsilon_d = -0.30$ in wind scenario 2 were analyzed.
If the number of operating thermal power plants is reduced by the demand response, the economic value of the wind farms will increase due to the high number of thermal power plants substituted by wind farms.
Figure \ref{fig:w2dr} (a) shows that the number of operating thermal power plants is reduced by the demand response during $9:00$ and $20:00$.
This implies that the effects of the demand response can be clearly observed in the number of operating thermal plants.
Figure \ref{fig:w2dr} (b) depicts the price structure with and without the demand response.
The rectangular structure for wind scenario 2 is smoothened by the demand response. 
The price during $17:00$ and $20:00$ is kept at a high level.
It is recognized that the marginal cost is high during $17:00$ and $20:00$ in Fig. \ref{fig:w2dr} (c).
Therefore, the high price during $17:00$ and $20:00$ is a consequence of the high marginal cost during the same period.
Because the operation cost was not increased in scenario 2, we cannot expect the demand response to cause a reduction in the operating cost in this case. 
However, a power utility has to continue to use thermal power plants for ensuring supply-demand balance; some of these plants can be decommissioned after installing a large number of wind farms or PV system,
if the demand response is applied using an appropriate price structure.

\section{Conclusion}

The output of renewable energy fluctuates significantly depending on weather conditions.
Therefore, it will be difficult to ensure supply-demand balance of electric power using currently used conventional power systems.
We have developed a unit commitment model to analyze requirements of the forecast output and its error for renewable energies.
Our model obtains the time series for the operational state of thermal power plants that would maximize the profits of an electric power utility by taking into account both the forecast of output and its error for renewable energies and the demand response of consumers.
The model is formulated as a mixed integer linear programming problem. 

We considered a power system consisting of thermal power plants, PV systems, and wind farms.
The unit commitment model was solved using a commercial solver to analyze the power system. 

First, the basic property of the model was discussed using the results of the reference case.
The fluctuation of the wind and the PV outputs was absorbed by starting the thermal power plants serially in the merit order.  
The price structure where the daytime rate was higher than the nighttime rate was obtained as a result of maximizing the profits of an electric power utility. 

Next, the effects of the forecast error on the operation cost and reserves were analyzed.
In addition to the reference case ($\sigma_w=3MW$), we analyzed two more cases, namely, $\sigma_w=\sigma_p=6MW$ and $\sigma_w=\sigma_p=9MW$.
We confirmed that the operation cost increased with the forecast error.
The intra-day structure of the spinning reserve was the same for the three cases, but the level increased with the forecast error. 
The LFC margin decreased during daytime, while it remained relatively high during nighttime, independently of the forecast error.

Then, the effects of a sudden decrease in wind power were analyzed.
More thermal power plants will have to be operated to generate power to absorb this sudden decrease in wind power.
The increase in the number of operating thermal power plants within a short period did not affect the total operation cost significantly;
however the substitution of thermal power plants by wind farms or PV systems is not expected to be very high.

Finally, the effects of the demand response in the case of a sudden decrease in wind power were analyzed.
We confirmed that the number of operating thermal power plants reduced by the demand response.
This implies that the number of operating thermal power plants is controlled efficiently by the demand response if an appropriate price structure is used. 
A power utility has to continue to use thermal power plants for ensuring supply-demand balance; some of these plants can be decommissioned after installing a large number of wind farms or PV systems,
if the demand response is applied using an appropriate price structure.

In future work, we intend to study the demand response technology and the forecast output of wind farms and PV systems 
using the developed unit commitment model for the maximizing the renewable energy integration in an actual power system.
We also plan to quantitatively study the issue of substitution of existing thermal power plants by renewable energy using a large set of relevant data \cite{Braun2008}.

\section*{Acknowledgment}
The authors would like to thank Prof. Kosuke KUROKAWA (Tokyo Institute of Technology) and Dr. Takashi OOZEKI (The National Institute of Advanced Industrial Science and Technology) for stimulating discussions and helpful comments.

\end{document}